# Population pulsation resonances of excitons in monolayer MoSe$_2$ with sub 1 µeV linewidth

**Authors**: John R. Schaibley[1], Todd Karin[1], Hongyi Yu[2], Jason S. Ross[3], Pasqual Rivera[1], Aaron M. Jones[1], Marie E. Scott[1], Jiaqiang Yan[4,5], D. G. Mandrus[4,5,6], Wang Yao[2], Kai-Mei Fu[1,7], Xiaodong Xu[1,3*]

Affiliations:
[1] Department of Physics, University of Washington, Seattle, Washington 98195, USA
[2] Department of Physics and Center of Theoretical and Computational Physics, University of Hong Kong, Hong Kong, China
[3] Department of Materials Science and Engineering, University of Washington, Seattle, Washington 98195, USA
[4] Materials Science and Technology Division, Oak Ridge National Laboratory, Oak Ridge, Tennessee, 37831, USA
[5] Department of Materials Science and Engineering, University of Tennessee, Knoxville, Tennessee, 37996, USA
[6] Department of Physics and Astronomy, University of Tennessee, Knoxville, Tennessee 37996, USA
[7] Department of Electrical Engineering, University of Washington, Seattle, Washington 98195, USA

*Correspondence to: xuxd@uw.edu

**Abstract:**

Monolayer transition metal dichalcogenides, a new class of atomically thin semiconductors, possess optically coupled 2D valley excitons. The nature of exciton relaxation in these systems is currently poorly understood. Here, we investigate exciton relaxation in monolayer MoSe$_2$ using polarization-resolved coherent nonlinear optical spectroscopy with high spectral resolution. We report strikingly narrow population pulsation resonances with two different characteristic linewidths of 1 µeV and <0.2 µeV at low-temperature. These linewidths are more than three orders of magnitude narrower than the photoluminescence and absorption linewidth, and indicate that a component of the exciton relaxation dynamics occurs on timescales longer than 1 ns. The ultra-narrow resonance (<0.2 µeV) emerges with increasing excitation intensity, and implies the existence of a long-lived state whose lifetime exceeds 6 ns.

**Text:**

The discovery of the direct bandgap nature and unique spin-valley coupled physics in monolayer transition metal dichalcogenides (TMDs) has sparked wide interest in understanding 2D valley excitons [1-3]. These electrically-tunable valley excitons allow for direct optical control of spin and valley degrees of freedom, promising for optoelectronics and valleytronics at the atomically thin limit. These applications hinge on the knowledge of key fundamental properties of valley excitons, such as the lifetimes and the intervalley scattering rate, which remain elusive due to the interplay between different excitonic states, inhomogeneous broadening, and many-body interaction effects. Recent progress towards understanding these 2D excitons includes: the optical generation of valley exciton polarization and quantum coherence [4-7], electrical tuning of excitonic charging effects [3,7,8], and the identification of exceptionally large exciton [9-15] and trion binding energies [3,8]. However, little is known about the nature of exciton relaxation in TMD monolayers beyond the ultra-fast timescale.

In a delocalized 2D excitonic system, the dispersion of the exciton center of mass momentum can be used to classify excitons relative to the photon dispersion, i.e., the light cone (Fig. 1(a)). Excitons inside of the light cone are referred to as bright excitons, since their center of mass momentum allows them to radiatively recombine. Conversely, dark excitons are located outside of the light cone, and cannot radiatively recombine until they scatter back into the light cone. Therefore, lifetimes for bright exciton are expected to be much shorter than for dark excitons. The presence of dark excitons can add a long lifetime component to the measured exciton decay time. Localized excitons might also exist (e.g. trap or defect-bound states) and could give rise to a long lifetime.

Recent low-temperature time-resolved measurements have shown that a fast component of exciton lifetime is on order of 1-100 ps [16-21] in a variety of TMD monolayer systems. The shortest of these lifetimes are within an order of magnitude of the theoretical predictions for the delocalized bright exciton radiative lifetime [21,22]; however, contributions from dark excitons are expected to significantly increase the measured exciton lifetime [22]. We note that the relatively low quantum yield (~0.1%) reported thus far in monolayer TMDs [1] appears to conflict with the interpretation that the intrinsic radiative decay is fast, and might imply that non-radiative decay processes dominate the ultra-fast exciton relaxation dynamics [23]. To probe these dynamics, techniques other than time resolved pump-probe and photoluminescence can be used to measure longer timescales which are often obscured in ultra-fast measurements.

In this Letter, we investigate the low-temperature relaxation rates of valley-excitons in monolayer $MoSe_2$ by performing high-resolution coherent nonlinear optical spectroscopy. We report evidence of exciton population dynamics that occur on the nanosecond timescale, which is 1-3 orders of magnitude longer than the lifetimes reported in ultra-fast time resolved measurements [16-21]. Our measurements reveal two distinct timescales: 1) a ~1-2 ns lifetime which is associated with the exciton decay time, 2) and a >6 ns lifetime which we assign to a previously unreported long-lived state. These time scales decrease as exciton density increases, which is evidence of strong many-body interaction effects in 2D semiconductors.

High-resolution coherent nonlinear optical spectroscopy has been used in atomic and solid state systems to probe relaxation processes which are obscured by inhomogeneous broadening and complex relaxation dynamics. In semiconductor optics, this technique was used extensively to study exciton relaxation [24], diffusion [25], and slow light effects [26] of GaAs excitons and nanocrystal systems [27].

We perform a continuous-wave two-color differential reflectivity (DR) measurement which is depicted in Fig. 1(b). Two continuously tunable, narrow bandwidth lasers are used: the pump laser is fixed on the exciton resonance while the probe laser is scanned through zero pump-probe detuning ($\Delta$) at high (<100 neV) resolution. The pump and probe lasers are amplitude modulated at two different frequencies, and a lock-in detection scheme is used to measure the nonlinear optical response at the difference of the modulation frequencies. The laser intensity is sufficiently low so that the dominant contribution to the non-degenerate DR signal is the third-order susceptibility ($\chi^{(3)}$) (Fig. 1(c)).

The data reported in the main text are from a single $MoSe_2$ monolayer on a $SiO_2$ substrate measured at 30 K. All experiments were performed in the reflection geometry with a beam diameter of about 1.5 µm. To characterize the sample, we first measure the degenerate DR response by scanning a single laser (split into pump and probe) in energy (Fig. 2(a)). Comparing the DR response to the photoluminescence spectrum (Fig. 2(a) inset), the DR resonances are assigned to the neutral and charged excitons at 1.655 eV and 1.625 eV, respectively [3]. This DR signal potentially contains contributions from phase-space filling, exciton-exciton interactions, bandgap renormalization by free carriers [28], and excitation-induced dephasing [29]. Optical interference between the sample and substrate [30] also contributes to the asymmetry of the DR lineshape by mixing the real and imaginary parts of the electric susceptibility.

In contrast to the few meV linewidth of the degenerate DR spectra, the high resolution two-color DR measurements show narrow (µeV) resonances, on top of a broad nonlinear signal which approximately follows the degenerate DR spectrum (Fig. 2(a)). Additional DR measurements on three other $MoSe_2$ monolayers show similar response [31]. Typical high resolution DR spectra are shown in Fig. 2(b-d) for cross-linearly polarized pump and probe at three different pump energy positions. Each DR spectrum is fit to a weighted sum of the real and imaginary parts of the same complex Lorentzian function ($\beta$) ($A$ Re($\beta$) + $B$ Im($\beta$), where the weights, A and B, are allowed to vary). The Lorentzian linewidth (FWHM) of the resonance is about 2 µeV. From the low to high energy side of the broad exciton resonance, the linewidth increases by approximately 25%, and an increase of the lineshape asymmetry is also observed. A qualitatively similar linewidth dependence has been observed in GaAs quantum well systems, explained by a phonon-assisted spectral diffusion process associated with exciton localization [25]. For the remainder of this work, high resolution DR measurements are performed near or below the exciton line center where the lineshape is nearly symmetric.

We examine pump power dependence of the resonance by fixing the pump laser near the peak of the exciton resonance (1.657 eV) and scanning the probe. Fig. 2(e) shows the linear dependence of linewidth as a function of pump power at probe power of 20 µW. Extrapolating to zero pump power, we obtain an intercept of 1.53 µeV, which is broadened by the probe power. At the lowest applied pump and probe power of 2 µW, the narrowest observed cross-linearly polarized DR resonance has a linewidth of 0.8 µeV (blue curve in Fig. 3(a)).

The DR responses show an interesting dependence on polarization (Figs. 3-4). In the first of these measurements, we compare the cross-linearly (blue) and cross-circularly (red) polarized pump and probe. The narrow resonance is negligible in the cross-circularly polarized DR response compared to the cross-linearly polarized response (Fig. 3(a)). At low power, the cross-linear, co-circular and co-linear polarized responses all have the narrow resonance dip with comparable linewidth (see Fig. 4 and the Supplementary Material for co-linear). Fig. 4 shows power dependent DR spectra for both cross-linearly and co-circularly polarized pump and probe. For cross-linearly polarized pump and probe (Fig. 4(a)), we observe a broadening of the linewidth with increasing power, consistent with Fig. 2(e), whereas the co-circularly (and co-linearly) polarized spectrum has a qualitatively different lineshape (Fig. 4(b)). Remarkably, for co-circularly polarized pump and probe, an extremely narrow (< 0.2 µeV FWHM) peak emerges from the dip with increasing laser intensity, eventually dominating the response (Fig. 4(b)).

In a simpler system, such as an inhomogeneously broadened ensemble of two-level atoms, the dominant DR response arises from the third-order susceptibility ($\chi^{(3)}$) when working in the low intensity limit. In a high-resolution pump-probe experiment, narrow resonances in $\chi^{(3)}$ can arise from both incoherent spectral hole-burning and coherent contributions from population pulsation [34,35] (Fig. 1(c)). Hole-burning resonances arise from the saturation of a narrow spectral distribution resulting in a decrease in probe absorption around the pump energy. The width of a hole-burning resonance provides a measure of the homogeneous spectral linewidth, including pure dephasing effects. Coherent population pulsation arises from the interference of the pump and probe laser fields through the excitation of the medium, leading to a modulation of the nonequilibrium populations at the pump-probe detuning. The population can follow the temporal modulation provided that the pump-probe detuning is smaller than the state's population decay rate. This coherent process leads to a resonance in the $\chi^{(3)}$ response whose spectral width provides a measure of the state lifetime [34]. Since population pulsation resonances are not directly sensitive to the dephasing rate, they can be orders of magnitude narrower than hole-burning resonances given because the dephasing rate is typically much larger than the population decay. Therefore population pulsation typically dominates the high resolution $\chi^{(3)}$ response [25]. In a system whose dephasing rate is much larger than a population decay rate, one expects a narrow population pulsation resonance on top of a broad nonlinear signal [25,34].

We attribute the narrow (~1 µeV) resonances in DR to coherent population pulsation due to both the narrow linewidths and their polarization dependence [31]. Whereas, the broad DR response likely arises from spectral hole-burning that is subject to strong spectral diffusion processes [36]. The nature of this broad nonlinear response will be explored in future work. Our assignments are further supported by recent independent measurements of the exciton homogeneous linewidth in TMD monolayers, which report the linewidth to be on the 1 meV-scale [21], three orders of magnitude larger than the ~1 µeV linewidth of the resonance reported here. We also note that quasi-2D excitons confined to GaAs quantum well structures also exhibited low temperature dephasing rates on the order of 0.5 meV and population pulsation resonance on the µeV-scale [25]. Thus, spectral-hole burning is unlikely to be the cause of the narrow resonances.

To understand the population pulsation resonances and their polarization dependence, we use the optical Bloch equations [37] to calculate the nonlinear susceptibility perturbatively in the $\chi^{(3)}$ limit [31]. In monolayer TMDs, we model the valley excitons in the +K and -K valleys as independent subsystems, which are coupled through intervalley relaxation (Fig. 3(b)). For each valley, we use a single level, $|\pm K\rangle$, to denote excitons both inside and outside the light cone (Fig. 3(b)) [31]. We note that this picture can be used as a model for weakly localized excitons. We also phenomenologically include a long-lived state $|ll\rangle$ to account for the extremely narrow linewidth (< 0.2 µeV) peak in the co-circularly polarized response. Possible candidates for such a state include the spin-forbidden or valley-forbidden excitons, and defect trapped states. The valley excitons can relax to such a long-lived state with a rate $\Gamma_{nr}$. In our model, a single rate $\Gamma_r$ accounts for the effective radiative decay rate of the exciton population inside and outside of the light cone [31]. The intervalley relaxation rate of excitons is denoted by $\Gamma_v$. The total decay rate of each valley exciton, $\Gamma_t$, is a sum of $\Gamma_v$, $\Gamma_r$, and $\Gamma_{nr}$. The non-radiative decay rate of the long-lived state is $\Gamma_{ll}$.

The polarization-dependent response can be understood by considering the optical selection rules for bright excitons in monolayer TMDs. Valley excitons in the +K and −K valleys are excited by circularly polarized light with the opposite helicity [38], as shown in Fig 3(b). Therefore, when both beams are cross-circularly polarized, there should be no population pulsation resonance since the pump and probe fields do not interfere when coupling to excitons in opposite valleys. This explains the negligible dip in the cross-circularly polarized case in Fig. 3(a). Even though the cross-circularly polarized configuration shows a negligible population pulsation response, the broad (~meV-scale) nonlinear response (away from zero detuning) is similar to the other polarizations, perhaps indicating the importance of intervalley scattering.

The existence of an ultra-narrow resonance which distinguishes the cross-linearly and co-circularly polarized DR responses (Fig. 4) can be explained by the optical Bloch equations using the energy model as depicted in Fig. 3(b); however, the sign change of the narrow peak relative to the broader dip requires more complicated theory which we explore in the Supplementary Material [31]. In the cross-linearly polarized configuration, the interference of the pump and probe fields only modulates the population difference between the +K and -K valley excitons, whereas the sum of the two populations is not modulated [31]. This results in a single Lorentzian in the cross-linearly polarized DR response of the form $\frac{1}{\Delta + i\,\Gamma}$, where $\Gamma = \Gamma_r + 2\Gamma_v + \Gamma_{nr}$. The pulsation effect from the long-lived state lies in the modulated sum population, and is therefore not present in the cross-linearly polarized measurement. The result of this analysis shows that the narrowest DR resonance in the cross-linearly polarized configuration provides a measure of the exciton average lifetime, which corresponds to the average decay rate $\Gamma_r$ of the bright and dark excitons [31]. An analysis of the role of dark excitonic states is explored in the Supplementary Material.

Assuming the usual relationship, $\Gamma^{-1} = 1/\pi\Delta\nu$, between lifetime ($\Gamma^{-1}$) and spectral width ($\Delta\nu$), holds for this system, we place a ~1.7 ns lower bound on the lifetime related to the cross-linearly polarized linewidth, $h\Delta\nu \sim 0.8\,\mu eV$ ($\Delta\nu = 0.19$ GHz) FWHM from Fig. 3(a). Using the model depicted in 3b, this rate places a lower bound on the overall exciton lifetime which is 1-3 orders of magnitude longer than the lifetimes recently reported in ultrafast pump-probe experiments [16-20]. We note that the 1.7 ns lifetime corresponds to an average of both the bright and dark exciton lifetimes and includes nonradiative decay to the long-lived state as well as intervalley scattering [31]. The observation of increasing linewidth as a function of power could indicate the influence of interaction effects [37,39] such as exciton-exciton annihilation [40] (See Fig. 2(e) and Fig. 4), where the power broadening corresponds to a decrease in the exciton lifetime.

When the pump and probe are co-circularly polarized, the fields only couple to bright excitons in one valley. In this case, the interference of the pump and probe fields modulates the exciton population in this valley. The population of the long-lived state is also modulated since it is populated by the relaxation from the $|\pm K\rangle$ exciton. This leads to two types of resonances. The first type is similar to the cross-linearly polarized resonance, whose linewidth is determined by $\Gamma_t$. The second type of resonance comes from the pulsation of the long-lived state, which contributes an additional resonance proportional to $\frac{\Gamma_{nr}}{(\Delta + i\Gamma_{ll})(\Delta + i(\Gamma_r + \Gamma_{nr}))}$. In the limit that the long-lived state decay is slow ($\Gamma_{ll} \ll \Gamma_r + \Gamma_{nr}$), the linewidth is given [41] by $\Gamma_{ll}$. A similar argument can be made for co-linearly polarized excitation [31].

The emergence of the ultra-narrow co-circular resonance with increasing intensity can be modeled by phenomenologically adding an exciton density dependence to the non-radiative decay rate, $\Gamma_{nr}$, which increases with exciton density (laser power). With this assumption, the contribution from the long-lived state is negligible at low power excitation. This is consistent with low power (10 μW) measurements showing similar (~2 μeV) linewidths for both cross-linear and co-circular polarizations, which measure $\Gamma_t$. However, under high power (40 μW) excitation, $\Gamma_{nr}$ increases, and the long-lived state contribution dominates the co-circularly polarized measurements. We note that recent transient absorption measurements, performed at room temperature on monolayer $MoS_2$, also indicate the importance of density dependent relaxation phenomena [40].

The narrow peak in the co-circular DR response indicates that the lifetime of the long-lived state is longer than 6.6 ns. The existence of this state, which appears to become more important at higher exciton densities, could be important for interpreting previous time-resolved measurements. For example, if decay to the long-lived state dominates valley exciton relaxation, time-resolved photoluminescence will be rapidly quenched and will effectively measure $\Gamma_{nr}$, whereas our high resolution coherent nonlinear spectroscopy technique is sensitive to long timescales and ground state dynamics. Relaxation to the long-lived state could also explain the low radiative quantum yield recently reported for monolayer TMDs [1], since the long-lived state can trap the exciton population. This model also gives insight into the recent observation of an increase of exciton lifetime with temperature, since at higher temperature, the long-lived state could scatter back to the bright exciton state by interacting with phonons [16].

**Acknowledgments:** We would like to acknowledge useful discussions with D.G. Steel, and H. Wang, and Xiaoqin Li. This work is mainly supported by the DoE BES (DE-SC0008145 and DE-SC0012509). AMJ is partially supported by NSF DGE-0718124. JR is partially supported by NSF DGE-1256082. TK and KF were supported by NSF DGE-1256082 and NSF 1150647. HY and WY are supported by the Croucher Foundation (Croucher Innovation Award), and the RGC of Hong Kong (HKU705513P, HKU9/CRF/13G). JY and DM are supported by US DoE, BES, Materials Sciences and Engineering Division. XX thanks the support from Cottrell Scholar Award. Device fabrication was performed at the University of Washington Nanofabrication Facility and Nanotech User Facility, both members of the NSF National Nanotechnology Infrastructure Network.
**Reference**

[1]  K. F. Mak, C. Lee, J. Hone, J. Shan, and T. F. Heinz, Phys Rev Lett **105** (2010).
[2]  A. Splendiani, L. Sun, Y. Zhang, T. Li, J. Kim, C.-Y. Chim, G. Galli, and F. Wang, Nano Lett. **10**, 1271 (2010).
[3]  J. S. Ross *et al.*, Nat Commun **4** (2013).
[4]  K. F. Mak, K. L. He, J. Shan, and T. F. Heinz, Nat Nanotechnol **7**, 494 (2012).
[5]  T. Cao *et al.*, Nat Commun **3** (2012).
[6]  H. Zeng, J. Dai, W. Yao, D. Xiao, and X. Cui, Nat Nano **7**, 490 (2012).
[7]  A. M. Jones *et al.*, Nat Nanotechnol **8**, 634 (2013).
[8]  K. F. Mak, K. L. He, C. Lee, G. H. Lee, J. Hone, T. F. Heinz, and J. Shan, Nat Mater **12**, 207 (2013).
[9]  C. Zhang, A. Johnson, C.-L. Hsu, L.-J. Li, and C.-K. Shih, Nano Lett. **14**, 2443 (2014).
[10] Z. Ye, T. Cao, K. O/'Brien, H. Zhu, X. Yin, Y. Wang, S. G. Louie, and X. Zhang, Nature **513**, 214 (2014).
[11] B. Zhu, X. Chen, and X. Cui, arXiv preprint arXiv:1403.5108 (2014).
[12] A. Chernikov, T. C. Berkelbach, H. M. Hill, A. Rigosi, Y. Li, O. B. Aslan, D. R. Reichman, M. S. Hybertsen, and T. F. Heinz, Phys Rev Lett **113**, 076802 (2014).
[13] K. He, N. Kumar, L. Zhao, Z. Wang, K. F. Mak, H. Zhao, and J. Shan, Phys Rev Lett **113**, 026803 (2014).
[14] G. Wang, X. Marie, I. Gerber, T. Amand, D. Lagarde, L. Bouet, M. Vidal, A. Balocchi, and B. Urbaszek, arXiv preprint arXiv:1404.0056 (2014).
[15] M. M. Ugeda *et al.*, Nat Mater **advance online publication** (2014).
[16] T. Korn, S. Heydrich, M. Hirmer, J. Schmutzler, and C. Schüller, Applied Physics Letters **99**, 102109 (2011).
[17] C. Mai, A. Barrette, Y. Yu, Y. Semenov, K. W. Kim, L. Cao, and K. Gundogdu, Nano Lett. **14**, 202 (2014).
[18] H. Shi, R. Yan, S. Bertolazzi, J. Brivio, B. Gao, A. Kis, D. Jena, H. G. Xing, and L. Huang, ACS nano **7**, 1072 (2013).
[19] D. Lagarde, L. Bouet, X. Marie, C. R. Zhu, B. L. Liu, T. Amand, P. H. Tan, and B. Urbaszek, Phys Rev Lett **112**, 047401 (2014).
[20] G. Wang, L. Bouet, D. Lagarde, M. Vidal, A. Balocchi, T. Amand, X. Marie, and B. Urbaszek, Phys Rev B **90**, 075413 (2014).
[21] G. Moody *et al.*, arXiv preprint arXiv:1410.3143 (2014).
11

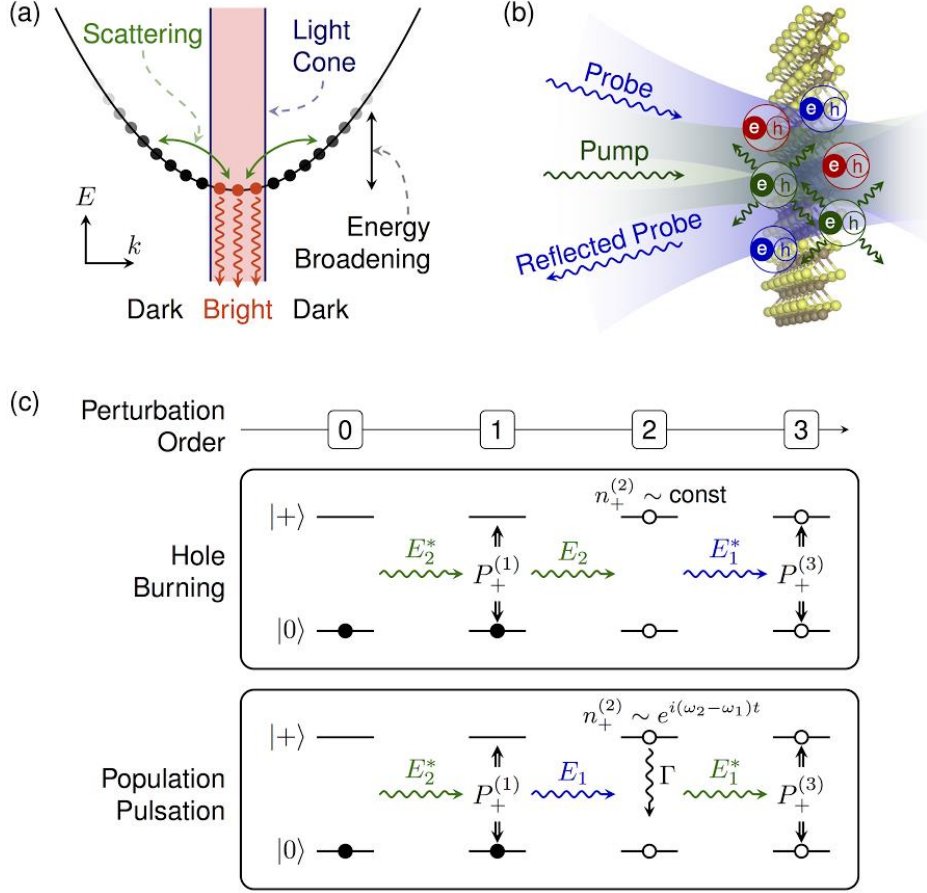

FIG. 1. (a) Depiction of the dispersion relation for delocalized excitons, where the x-axis represents the exciton center of mass momentum. In this picture, only excitons inside of the light cone can efficiently radiatively recombine. (b) Depiction of pump and probe fields interacting with a MoSe$_2$ monolayer. A narrow bandwidth pump laser selectively excites excitons of a particular energy (shown here as green). The interference of the pump and probe laser fields gives rise to a population pulsation resonance. (c) In an inhomogeneous distribution of two-level systems composed of states ($|0\rangle$ and $|+\rangle$), both hole-burning and population pulsation effects can contribute to a $\chi^{(3)}$ response. Here, we show two examples of the perturbation up to third order in the applied fields. $E_1$ and $E_2$ are the probe and pump fields at frequencies $\omega_1$ and $\omega_2$ respectively. $P_+$ is the polarization, and $n_+$ is the excited state population. The superscripts label the perturbation order. The population pulsation effect results in a modulation of the state populations at the pump-probe difference frequency.

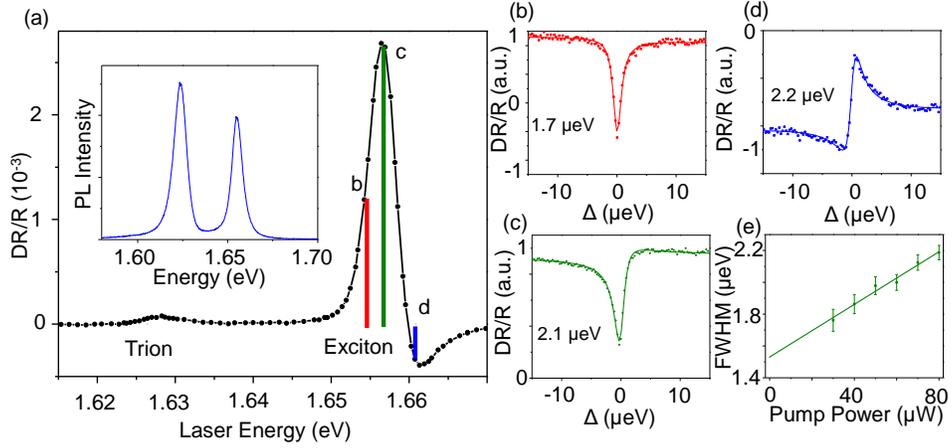

FIG. 2. (a) Degenerate DR spectrum showing the exciton and trion resonance, consistent with low temperature photoluminescence (PL) measurements (inset). The colored lines indicate the spectral positions of the non-degenerate DR spectra shown on the right. (b-d) High resolution non-degenerate DR measurements as a function of pump position (as indicated in inset a), showing narrow DR resonances for different pump energies (cross-linearly polarized pump and probe). Linewidths are extracted from fits of a weighted contribution from both the real and imaginary parts of the same complex Lorentzian function. (e) The linewidth is observed to increase linearly with increasing pump power revealing a zero power intercept of $1.53 \pm 0.04$ µeV and a slope of $8.3 \pm 0.6$ neV/µW (for 20 µW probe power at pump position c).

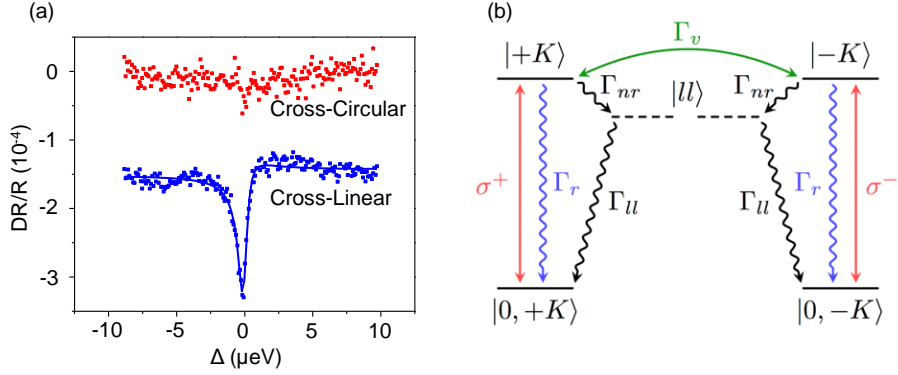

FIG. 3. Polarization dependent differential reflectivity. (a) The low power (2 µW pump and probe) DR response has a linewidth of 0.8 µeV FWHM for cross-linearly polarized pump and probe (blue). The negligible resonance for cross-circularly polarized pump and probe (red) is a consequence of the valley-dependent optical selection rules. The pump energy is 1.655 eV. (b) Energy level diagram of the valley exciton system. Direct exciton transitions occur at K-valleys forming two exciton subsystems with $\sigma \pm$ polarized optical selection rules. We model the exciton as a single state for each valley. The exciton relaxation rate ($\Gamma_r$), intervalley relaxation rate ($\Gamma_v$), nonradiative decay rate to long-lived state ($\Gamma_{nr}$) and long-lived state decay rate ($\Gamma_{ll}$) are shown.

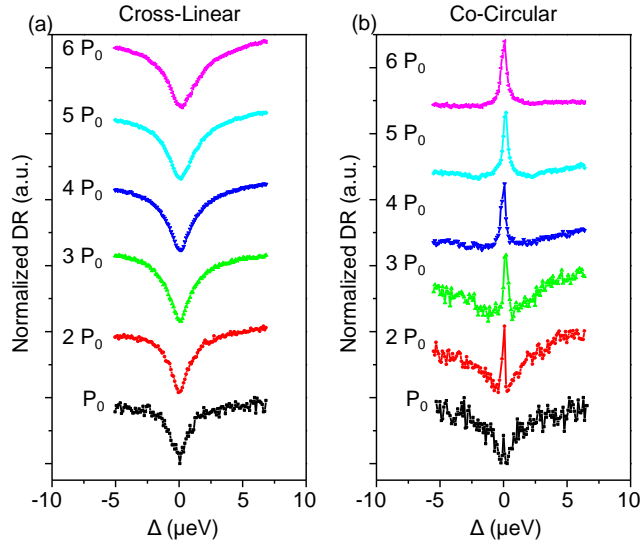

FIG. 4. Power dependent cross-linear and co-circular differential reflectivity. The reflected pump and probe are simultaneously detected and the two powers are increased together. $P_0$ corresponds to 10 µW. At $P_0$, the cross-linearly (a) and co-circularly (b) polarized spectra have similar linewidths of 1.8 µeV and 2.5 µeV respectively. The amplitude of each spectrum is normalized, and the data are stacked to highlight the changes of lineshape with power. The cross-linearly polarized data show a power dependent broadening, consistent with an increase of the nonradiative decay rate, $\Gamma_{nr}$. At higher powers a narrow peak emerges in the co-circularly polarized spectra whose width is related to the decay rate of the long-lived states, $\Gamma_{ll}$.

**Supplemental Material**

**Population pulsation resonances of excitons in monolayer MoSe$_2$ with sub 1 μeV linewidth**

John R. Schaibley, Todd Karin, Hongyi Yu, Jason S. Ross, Pasqual Rivera, Aaron M. Jones, Marie E. Scott, Jiaqiang Yan, D. G. Mandrus, Wang Yao, Kai-Mei Fu, Xiaodong Xu

## Supplementary Figures

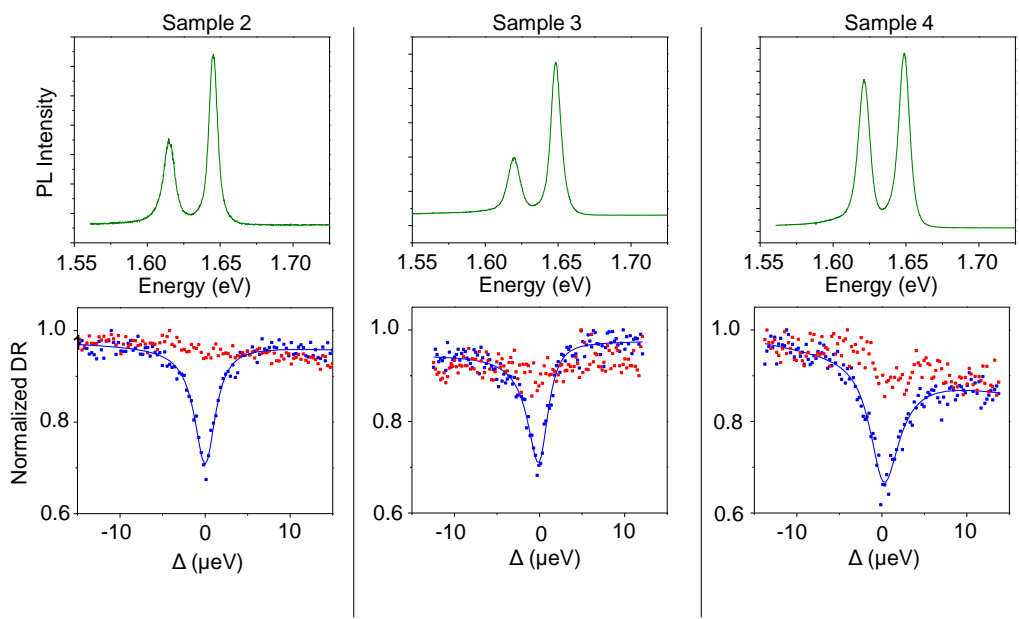

FIG. S1. DR measurements on additional samples. Low temperature PL (green) and cross-linearly(blue)/circularly(red) polarized DR measurements are performed on three different MoSe$_2$ monolayers. Pump and probe powers are both 40 μW. The cross-linearly polarized DR linewidth for sample 2, 3, 4 is 3.0 μeV, 2.9 μeV, and 4.2 μeV FWHM respectively. The DR spectra are normalized so that the maximum is equal to one.

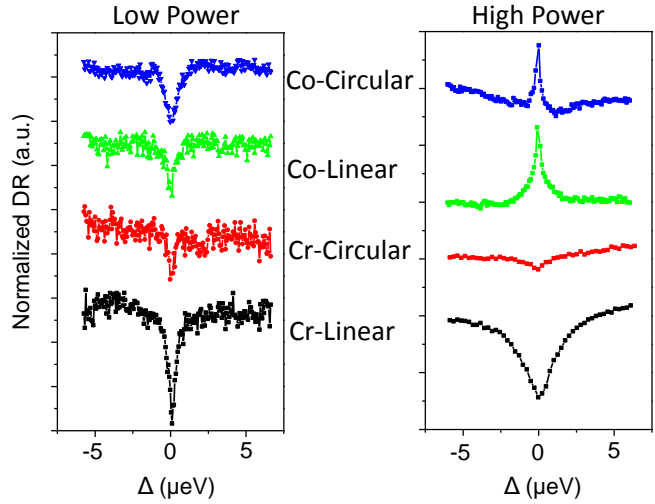

FIG. S2. Polarization dependent DR. Low and high power polarization dependent DR for the four combinations of co/cross-linearly/circularly polarized pump and probe. At low power (2 µW), we observe DR dips revealing linewidths of ~1 µeV FWHM. The weak cross-circular resonance is consistent with the slight ellipticity of the pump and probe fields. At high power (40 µW), the cross-linear linewidth broadens, and the cross-circular signal remains weak; however, the co-linear and co-circular signals show a narrow ~0.2 µeV resonance peak. The pump energy is 1.655 eV.

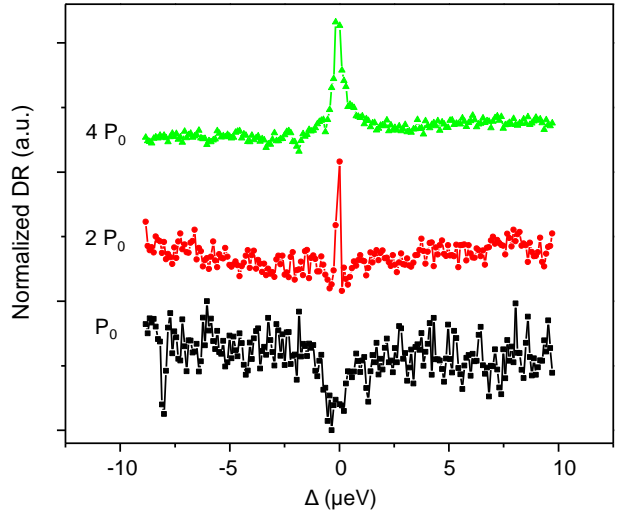

FIG. S3. Power dependence of the co-linear DR signal. High resolution non-degenerate DR with co-linearly polarized pump and probe. The pump and probe powers are equal and are both detected. $P_0$ corresponds to 5 µW. The pump energy is 1.655 eV.

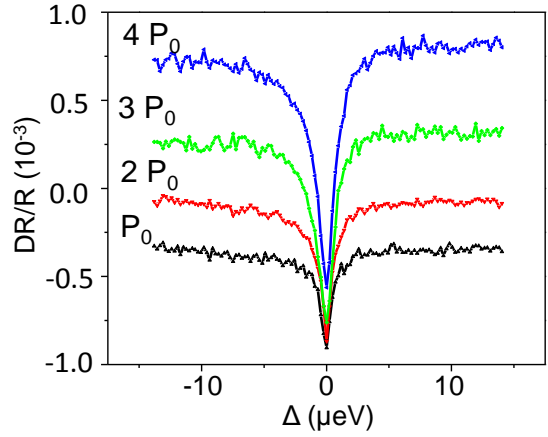

FIG. S4. Power dependence of the cross-linear DR signal at low power. High resolution non-degenerate DR with cross-linearly polarized pump and probe. The nonlinear offset signal changes sign with power, but the population pulsation dip is always negative, corresponding to reduced reflection. The probe power is 1 µW, and the pump power is varied. $P_0$ corresponds to 10 µW. The pump energy is ~1.655 eV.

|  | Decay Rate | Linewidth (µeV) | Lifetime (ns) |
|---|---|---|---|
| Exciton | $\Gamma = \Gamma_r + 2\Gamma_v + \Gamma_{nr}$ | <0.8 | >1.7 |
| Long-Lived State | $\Gamma_{ll}$ | <0.2 | >6.6 |

Table S1. Summary of linewidths and lifetimes extracted from the population pulsation resonances with definitions given in the main text.

**Supplementary Note 1: Optical susceptibility equations in the two 2-level picture**

The obtained probe differential reflectance signal is proportional to $A \cdot \text{Re}(\Delta\chi(\omega)) + B \cdot \text{Im}(\Delta\chi(\omega))$ (Ref. [1]). Here $\Delta\chi(\omega)$ is the difference of the monolayer $MoSe_2$ optical susceptibility $\chi(\omega)$ with and without the pump laser, and is a function of laser frequency $\omega$. $A$ and $B$ are real numbers determined by the thickness and refractive index of the substrates, and are approximated as constants in the relevant frequency range of this experiment.

In this section, we perturbatively solve the optical susceptibility up to the third order of the pump laser field ($\Delta\chi(\omega) \approx \chi^{(3)}(\omega)$), using a simple picture of two 2-level systems plus long-lived states $|ll\rangle$ (see Fig. 3b of the maintext). When compared to the solutions of the excitonic Bloch equations (given in Supplementary Note 2), we find their difference is only quantitative. We show this simple picture can already explain most of the experimental observations, and gives the correct results of population decay rates.

The system dynamics can be separated into two parts: $d\hat{\rho}/dt = \frac{d\hat{\rho}}{dt}\big|_c + \frac{d\hat{\rho}}{dt}\big|_{inc}$, where $\hat{\rho}$ is the full density matrix for the system. Here, $\frac{d\hat{\rho}}{dt}\big|_c$ is the evolution under the Hamiltonian that describes the coherent field-exciton interaction, which acts as the generating source of exciton density and coherence. $\frac{d\hat{\rho}}{dt}\big|_{inc}$ is the incoherent process which describes the decay. The long-lived states only affect $\frac{d\hat{\rho}}{dt}\big|_{inc}$.

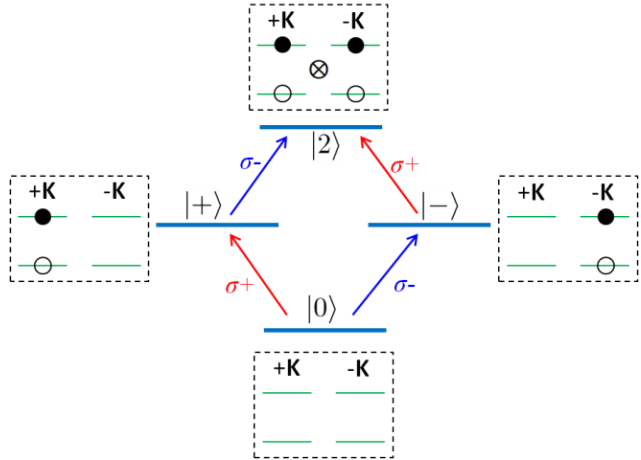

FIG. S5. Level diagram for coherent processes. In order to calculate the response to third order in the applied laser fields, the four states of the two 2-level systems and the optical transition selection rules between them are shown. We do not consider biexciton states here, assuming that the biexciton binding energy, on the order of 10 meV[2,3] is outside the range of pump-probe detuning of 10 µeV.

First, we consider $\frac{d\hat{\rho}}{dt}\big|_c$. The two 2-level systems can be re-expressed by the level diagram shown in Fig. S5, which contains a vacuum state $|0\rangle$ with no exciton, a $\pm K$ valley one-exciton state $|\pm\rangle$ and a two-exciton state $|2\rangle$. The Hamiltonian for such a system is:

$$\hat{H} = \omega_X(|+\rangle\langle+| + |-\rangle\langle-|) + 2\omega_X|2\rangle\langle 2| - \Omega_+(|+\rangle\langle 0| + |2\rangle\langle-|) \\ - \Omega_+^*(|0\rangle\langle+| + |-\rangle\langle 2|) - \Omega_-(|-\rangle\langle 0| + |2\rangle\langle+|) - \Omega_-^*(|0\rangle\langle-| + |+\rangle\langle 2|). \tag{1}$$

Here $\Omega_\pm$ is the $\hat{\sigma}_\pm$ component of the total Rabi frequency of the applied lasers, and $\omega_X$ is the exciton energy. We have taken $\hbar$ equal to 1 for simplicity. The equations of motion $i\frac{d\hat{\rho}}{dt}\big|_c \equiv [\hat{H}, \hat{\rho}]$ give

$$\begin{aligned}
i\frac{d\rho_{+0}}{dt}\bigg|_c &= \omega_X \rho_{+0} - (\rho_{00} - \rho_{++})\Omega_+ - \rho_{20}\Omega_-^* + \rho_{+-}\Omega_-, \\
i\frac{d\rho_{2-}}{dt}\bigg|_c &= \omega_X \rho_{2-} - (\rho_{--} - \rho_{22})\Omega_+ + \rho_{20}\Omega_-^* - \rho_{+-}\Omega_-, \\
i\frac{d\rho_{-0}}{dt}\bigg|_c &= \omega_X \rho_{-0} - (\rho_{00} - \rho_{--})\Omega_- - \rho_{20}\Omega_+^* + \rho_{-+}\Omega_+, \\
i\frac{d\rho_{2+}}{dt}\bigg|_c &= \omega_X \rho_{2+} - (\rho_{++} - \rho_{22})\Omega_- + \rho_{20}\Omega_+^* - \rho_{-+}\Omega_+, \\
i\frac{d\rho_{20}}{dt}\bigg|_c &= 2\omega_X \rho_{20} - (\rho_{-0} - \rho_{2+})\Omega_+ - (\rho_{+0} - \rho_{2-})\Omega_-, \\
i\frac{d\rho_{+-}}{dt}\bigg|_c &= -(\rho_{0-} - \rho_{+2})\Omega_+ - (\rho_{2-} - \rho_{+0})\Omega_-^*, \\
\frac{d\rho_{00}}{dt}\bigg|_c &= 2\text{Im}[\Omega_+ \rho_{0+}] + 2\text{Im}[\Omega_- \rho_{0-}], \\
\frac{d\rho_{++}}{dt}\bigg|_c &= -2\text{Im}[\Omega_+ \rho_{0+}] + 2\text{Im}[\Omega_- \rho_{+2}], \\
\frac{d\rho_{--}}{dt}\bigg|_c &= -2\text{Im}[\Omega_- \rho_{0-}] + 2\text{Im}[\Omega_+ \rho_{-2}], \\
\frac{d\rho_{22}}{dt}\bigg|_c &= -2\text{Im}[\Omega_- \rho_{+2}] - 2\text{Im}[\Omega_+ \rho_{-2}],
\end{aligned} \tag{2}$$

Assuming $\Omega$ is weak, we expand the density matrix in powers of the applied field, such that $\rho \approx \rho^{(0)} + \rho^{(1)} + \rho^{(2)} + \cdots$. To zeroth order in the applied fields, the only nonzero element is $\rho_{00}^{(0)} = 1$. Up to the 3rd order in the applied fields, $\rho_{00} \approx 1 + \rho_{00}^{(2)}$, $\rho_{\pm\pm} \approx \rho_{\pm\pm}^{(2)}$, $\rho_{22} \approx 0$, $\rho_{\pm 0} \approx \rho_{\pm 0}^{(1)} + \rho_{\pm 0}^{(3)}$, $\rho_{2\pm} \approx \rho_{2\pm}^{(3)}$, $\rho_{20} \approx \rho_{20}^{(2)}$, $\rho_{+-} \approx \rho_{+-}^{(2)}$.

Here we are mainly interested in the polarization $P_\pm = \rho_{\pm 0} + \rho_{2\mp}$, and the exciton population in each valley $n_\pm = \rho_{\pm\pm} + \rho_{22}$. They satisfy $i\frac{dP_\pm}{dt}\big|_c = \omega_X P_\pm - (\rho_{00} - \rho_{22} \mp (n_+ - n_-))\Omega_\pm$ and $\frac{dn_\pm}{dt}\big|_c = -2\text{Im}[\Omega_\pm P_\pm^*]$. By writing $P_\pm^{(1)} = \rho_{\pm 0}^{(1)}$, $P_\pm^{(3)} = \rho_{\pm 0}^{(3)} + \rho_{2\pm}^{(3)}$ and $n_\pm^{(2)} = \rho_{\pm\pm}^{(2)}$, they can be expressed as

$$i\frac{dP^{(1)}_{\pm}}{dt}\bigg|_c = \omega_X P^{(1)}_{\pm} - \Omega_{\pm},$$

$$\frac{dn^{(2)}_{\pm}}{dt}\bigg|_c = -2\text{Im}\left[\Omega_{\pm}\left(P^{(1)}_{\pm}\right)^*\right],$$

$$\frac{d\rho^{(2)}_{00}}{dt}\bigg|_c = 2\text{Im}\left[\Omega_+\left(P^{(1)}_+\right)^*\right] + 2\text{Im}\left[\Omega_-\left(P^{(1)}_-\right)^*\right], \quad (3)$$

$$i\frac{dP^{(3)}_{\pm}}{dt}\bigg|_c = \omega_X P^{(3)}_{\pm} - \left(\rho^{(2)}_{00} \mp \left(n^{(2)}_+ - n^{(2)}_-\right)\right)\Omega_{\pm}.$$

When linearly polarized lasers are applied, we rewrite the equations of motion in a linear basis, using $\Omega_x \equiv \frac{\Omega_+ + \Omega_-}{\sqrt{2}}$, $\Omega_y \equiv -i\frac{\Omega_+ - \Omega_-}{\sqrt{2}}$, $P_x \equiv \frac{P_+ + P_-}{\sqrt{2}}$ and $P_y \equiv -i\frac{P_+ - P_-}{\sqrt{2}}$. We also introduce the total exciton population $N \equiv n_+ + n_-$, and population difference of excitons in two valleys $\Delta N \equiv n_+ - n_-$, resulting in

$$i\frac{dP^{(1)}_{x/y}}{dt}\bigg|_c = \omega_X P^{(1)}_{x/y} - \Omega_{x/y},$$

$$i\frac{dP^{(3)}_x}{dt}\bigg|_c = \omega_X P^{(3)}_x - \rho^{(2)}_{00}\Omega_x + i\Delta N^{(2)}\Omega_y, \quad (4)$$

$$i\frac{dP^{(3)}_y}{dt}\bigg|_c = \omega_X P^{(3)}_y - \rho^{(2)}_{00}\Omega_y - i\Delta N^{(2)}\Omega_x.$$

The incoherent (decay) processes, $\frac{d\hat{\rho}}{dt}\big|_{\text{inc}}$, are described by the Lindblad master equation,

$$\frac{d\hat{\rho}}{dt}\bigg|_{\text{inc}} = \sum_m \left(L_m\hat{\rho}L^\dagger_m - \tfrac{1}{2}L^\dagger_m L_m\hat{\rho} - \tfrac{1}{2}\hat{\rho}L^\dagger_m L_m\right), \quad (5)$$

where $L_m$ are the Lindblad operators representing: radiative recombination, $L_1 = \sqrt{\Gamma_r}|0\rangle\langle+|$, $L_2 = \sqrt{\Gamma_r}|0\rangle\langle-|$, $L_3 = \sqrt{\Gamma_r}|+\rangle\langle 2|$, $L_4 = \sqrt{\Gamma_r}|-\rangle\langle 2|$; exciton homogeneous dephasing, $L_5 = \sqrt{2\gamma_h}(|+\rangle\langle+| + |-\rangle\langle-|)$; valley dephasing, $L_6 = \sqrt{2\gamma_v}(|+\rangle\langle+| - |-\rangle\langle-|)$; valley relaxation, $L_7 = \sqrt{\Gamma_v}|-\rangle\langle+|$, $L_8 = \sqrt{\Gamma_v}|+\rangle\langle-|$. The state $|ll\rangle$ brings additional decay channels which are denoted by $L_9 = \sqrt{\Gamma_{\text{nr}}}|ll\rangle\langle+|$, $L_{10} = \sqrt{\Gamma_{\text{nr}}}|ll\rangle\langle-|$ and $L_{11} = \sqrt{\Gamma_{ll}}|0\rangle\langle ll|$. Note that because $\rho_{22} = 0$ up to the third order, we have ignored all 2-exciton states and the corresponding decay process.

From the master equation we obtain,

$$\left.\frac{dP_\pm}{dt}\right|_{\text{inc}} = -\gamma P_\pm, \quad \left(\gamma = \frac{\Gamma_r + \Gamma_v + \Gamma_{\text{nr}}}{2} + \gamma_h + \gamma_v\right)$$

$$\left.\frac{dn_\pm}{dt}\right|_{\text{inc}} = -(\Gamma_r + \Gamma_v + \Gamma_{\text{nr}})n_\pm + \Gamma_v n_\mp,$$

$$\left.\frac{dn_{ll}}{dt}\right|_{\text{inc}} = -\Gamma_{ll} n_{ll} + \Gamma_{\text{nr}}(\rho_{++} + \rho_{--}),$$

$$\left.\frac{d\rho_{00}}{dt}\right|_{\text{inc}} = \Gamma_r(\rho_{++} + \rho_{--}) + \Gamma_{ll} n_{ll}.$$
(6)

Here, $\gamma$ is the homogeneous spectral linewidth, and $n_{ll}$ is the total population of all long-lived states. Using $\frac{d\hat{\rho}}{dt} = \left.\frac{d\hat{\rho}}{dt}\right|_c + \left.\frac{d\hat{\rho}}{dt}\right|_{\text{inc}}$ and $\rho_{00}^{(2)} + n_+^{(2)} + n_-^{(2)} + n_{ll}^{(2)} = 0$, we solve for the density matrix elements iteratively:

$$i\frac{dP_\pm^{(1)}}{dt} = (\omega_X - i\gamma)P_\pm^{(1)} - \Omega_\pm,$$

$$\frac{dn_\pm^{(2)}}{dt} = -2\text{Im}\left[\Omega_\pm \left(P_\pm^{(1)}\right)^*\right] - (\Gamma_r + \Gamma_v + \Gamma_{\text{nr}})n_\pm^{(2)} + \Gamma_v n_\mp^{(2)},$$

$$\frac{dn_{ll}^{(2)}}{dt} = -\Gamma_{ll} n_{ll}^{(2)} + \Gamma_{\text{nr}}\left(n_+^{(2)} + n_-^{(2)}\right),$$

$$i\frac{dP_+^{(3)}}{dt} = (\omega_X - i\gamma)P_+^{(3)} - \left(\rho_{00}^{(2)} - n_+^{(2)} + n_-^{(2)}\right)\Omega_+$$
$$= (\omega_X - i\gamma)P_+^{(3)} + \left(n_{ll}^{(2)} + 2n_+^{(2)}\right)\Omega_+,$$

$$i\frac{dP_-^{(3)}}{dt} = (\omega_X - i\gamma)P_-^{(3)} - \left(\rho_{00}^{(2)} + n_+^{(2)} - n_-^{(2)}\right)\Omega_-$$
$$= (\omega_X - i\gamma)P_-^{(3)} + \left(n_{ll}^{(2)} + 2n_-^{(2)}\right)\Omega_-.$$
(7)

For linearly polarized lasers, the above equations can be written as,

$$i\frac{dP_{x,y}^{(1)}}{dt} = (\omega_X - i\gamma)P_{x,y}^{(1)} - \Omega_{x,y},$$

$$\frac{dN^{(2)}}{dt} = -2\text{Im}\left[\Omega_x\left(P_x^{(1)}\right)^* + \Omega_y\left(P_y^{(1)}\right)^*\right] - (\Gamma_r + \Gamma_{\text{nr}})N^{(2)},$$

$$\frac{d\Delta N^{(2)}}{dt} = -2\text{Im}\left[-i\Omega_x\left(P_y^{(1)}\right)^* + i\Omega_y\left(P_x^{(1)}\right)^*\right] - (\Gamma_r + 2\Gamma_v + \Gamma_{\text{nr}})\Delta N^{(2)},$$

$$\frac{dn_{ll}^{(2)}}{dt} = -\Gamma_{ll} n_{ll}^{(2)} + \Gamma_{\text{nr}} N^{(2)},$$

$$i\frac{dP_x^{(3)}}{dt} = (\omega_X - i\gamma)P_x^{(3)} + \left(n_{ll}^{(2)} + N^{(2)}\right)\Omega_x + i\Delta N^{(2)}\Omega_y,$$

$$i\frac{dP_y^{(3)}}{dt} = (\omega_X - i\gamma)P_y^{(3)} + \left(n_{ll}^{(2)} + N^{(2)}\right)\Omega_y - i\Delta N^{(2)}\Omega_x.$$
(8)

We now consider various pump-probe polarization combinations. The pump (probe) laser has a Rabi frequency of $\Omega_p$ ($\Omega_b$) and an optical frequency $\omega_p$ ($\omega_b$). We denote $\Delta_p \equiv \omega_p - \omega_X$ the pump detuning and $\Delta \equiv \omega_b - \omega_p$ the probe-pump detuning. By picking out the $\omega_b$ frequency component $P^{(3)}(\omega_b)$ in $P^{(3)}$, we get the probe nonlinear susceptibility $\chi^{(3)}(\omega_b) = \frac{P^{(3)}(\omega_b)}{\Omega_b e^{-i\omega_b t}}$.

In the co-circular case, $\Omega_+ = \Omega_p e^{-i\omega_p t} + \Omega_b e^{-i\omega_b t}$ and $\Omega_- = 0$, so $P_+^{(1)} = -\frac{\Omega_p e^{-i\omega_p t}}{\Delta_p + i\gamma} - \frac{\Omega_b e^{-i\omega_b t}}{\Delta + \Delta_p + i\gamma}$, $P_-^{(1)} = 0$. The interference between the pump and the probe lasers leads to exciton population pulsation with frequency $\Delta$: $N^{(2)} = \frac{2\gamma}{\Gamma_r + \Gamma_{nr}} \left( \frac{|\Omega_p|^2}{\Delta_p^2 + \gamma^2} + \frac{|\Omega_b|^2}{(\Delta + \Delta_p)^2 + \gamma^2} \right) + \left( A e^{-i\Delta t} + \text{c.c.} \right)$, $\Delta N^{(2)} = \frac{2\gamma}{\Gamma_r + 2\Gamma_v + \Gamma_{nr}} \left( \frac{|\Omega_p|^2}{\Delta_p^2 + \gamma^2} + \frac{|\Omega_b|^2}{(\Delta + \Delta_p)^2 + \gamma^2} \right) + \left( B e^{-i\Delta t} + \text{c.c.} \right)$, with $A = \frac{\Omega_b \Omega_p^*}{\Delta + i(\Gamma_r + \Gamma_{nr})} \frac{\Delta + 2i\gamma}{(\Delta_p - i\gamma)(\Delta + \Delta_p + i\gamma)}$ and $B = \frac{\Omega_b \Omega_p^*}{\Delta + i(\Gamma_r + 2\Gamma_v + \Gamma_{nr})} \frac{\Delta + 2i\gamma}{(\Delta_p - i\gamma)(\Delta + \Delta_p + i\gamma)}$. The long-lived state inherits the population pulsation from the total exciton population $N^{(2)}$: $n_{ll}^{(2)} = \frac{2\Gamma_{nr}\gamma}{\Gamma_{ll}(\Gamma_r + \Gamma_{nr})} \left( \frac{|\Omega_p|^2}{\Delta_p^2 + \gamma^2} + \frac{|\Omega_b|^2}{(\Delta + \Delta_p)^2 + \gamma^2} \right) + \left( \frac{i\Gamma_{nr}}{\Delta + i\Gamma_{ll}} A e^{-i\Delta t} + \text{c.c.} \right)$. The optical susceptibility of the probe field is,

$$\chi_{\text{co-cir}}^{(3)}(\omega_b) = \alpha \frac{2\gamma}{\Delta + \Delta_p + i\gamma} \left( \frac{|\Omega_p|^2}{\Delta_p^2 + \gamma^2} + \frac{|\Omega_b|^2}{(\Delta + \Delta_p)^2 + \gamma^2} \right) + \frac{|\Omega_p|^2 (\Delta + 2i\gamma)}{(\Delta + \Delta_p + i\gamma)^2 (\Delta_p - i\gamma)} \left( \frac{1 + \frac{i\Gamma_{nr}}{\Delta + i\Gamma_D}}{\Delta + i(\Gamma_r + \Gamma_{nr})} + \frac{1}{\Delta + i(\Gamma_r + 2\Gamma_v + \Gamma_{nr})} \right). \tag{9}$$

Where $\alpha \equiv \frac{\Gamma_{nr}}{\Gamma_{ll}(\Gamma_r + \Gamma_{nr})} + \frac{1}{\Gamma_r + \Gamma_{nr}} + \frac{1}{\Gamma_r + 2\Gamma_v + \Gamma_{nr}}$ is a constant.

In the cross-circular case, $\Omega_+ = \Omega_p e^{-i\omega_p t}$ and $\Omega_- = \Omega_b e^{-i\omega_b t}$. The pump and probe lasers do not interfere so there is no population pulsation. The susceptibility of the probe field is,

$$\chi_{\text{cross-cir}}^{(3)}(\omega_b) = \frac{2\gamma}{\Delta + \Delta_p + i\gamma} \left( \frac{\alpha |\Omega_p|^2}{\Delta_p^2 + \gamma^2} + \frac{\beta |\Omega_b|^2}{(\Delta + \Delta_p)^2 + \gamma^2} \right). \tag{10}$$

Where $\alpha = \frac{\Gamma_{ll} + \Gamma_{nr}}{\Gamma_{ll}(\Gamma_r + \Gamma_{nr})} - \frac{1}{\Gamma_r + 2\Gamma_v + \Gamma_{nr}}$ and $\beta = \frac{\Gamma_{ll} + \Gamma_{nr}}{\Gamma_{ll}(\Gamma_r + \Gamma_{nr})} + \frac{1}{\Gamma_r + 2\Gamma_v + \Gamma_{nr}}$ are both constants.

In the co-linear case, $\Omega_x = \Omega_p e^{-i\omega_p t} + \Omega_b e^{-i\omega_b t}$ and $\Omega_y = 0$, so $P_x^{(1)} = -\frac{\Omega_p e^{-i\omega_p t}}{\Delta_p + i\gamma} - \frac{\Omega_b e^{-i\omega_b t}}{\Delta + \Delta_p + i\gamma}$, $P_y^{(1)} = 0$, $\Delta N^{(2)} = 0$. Now there are pulsations for the total exciton population $N^{(2)} = \frac{2\gamma}{\Gamma_r + \Gamma_{nr}} \left( \frac{|\Omega_p|^2}{\Delta_p^2 + \gamma^2} + \frac{|\Omega_b|^2}{(\Delta + \Delta_p)^2 + \gamma^2} \right) + \left( A e^{-i\Delta t} + \text{c.c.} \right)$ and the long-lived state population $n_{ll}^{(2)} =$

$$\frac{2\Gamma_{nr}\gamma}{\Gamma_{ll}(\Gamma_r+\Gamma_{nr})}\left(\frac{|\Omega_p|^2}{\Delta_p^2+\gamma^2}+\frac{|\Omega_b|^2}{(\Delta+\Delta_p)^2+\gamma^2}\right)+\left(\frac{i\Gamma_{nr}}{\Delta+i\Gamma_{ll}}Ae^{-i\Delta t}+\text{c.c.}\right).\text{ Here }A=\frac{\Omega_b\Omega_p^*}{\Delta+i(\Gamma_r+\Gamma_{nr})}\frac{\Delta+2i\gamma}{(\Delta_p-i\gamma)(\Delta+\Delta_p+i\gamma)}.$$

The susceptibility of the probe field is,

$$\chi^{(3)}_{\text{co-lin}}(\omega_b)=\alpha\frac{2\gamma}{\Delta+\Delta_p+i\gamma}\left(\frac{|\Omega_p|^2}{\Delta_p^2+\gamma^2}+\frac{|\Omega_b|^2}{(\Delta+\Delta_p)^2+\gamma^2}\right)\\+\frac{|\Omega_p|^2(\Delta+2i\gamma)}{(\Delta+\Delta_p+i\gamma)^2(\Delta_p-i\gamma)}\frac{1+\frac{i\Gamma_{nr}}{\Delta+i\Gamma_{ll}}}{\Delta+i(\Gamma_r+\Gamma_{nr})}, \quad (11)$$

with $\alpha\equiv\frac{\Gamma_{nr}}{\Gamma_{ll}(\Gamma_r+\Gamma_{nr})}+\frac{1}{\Gamma_r+\Gamma_{nr}}$.

In the cross-linear case, $\Omega_x=\Omega_p e^{-i\omega_p t}$ and $\Omega_y=\Omega_b e^{-i\omega_b t}$. The pump-probe interference leads to a pulsation of the population difference between the two excitons $\Delta N^{(2)}=\frac{i\Omega_b\Omega_p^*}{\Delta+i(\Gamma_r+2\Gamma_v+\Gamma_{nr})}\frac{\Delta+2i\gamma}{(\Delta_p-i\gamma)(\Delta+\Delta_p+i\gamma)}e^{-i\Delta t}+\text{c.c.}$. There is no pulsation in the total exciton population $N^{(2)}=\frac{2\gamma}{\Gamma_r+\Gamma_{nr}}\left(\frac{|\Omega_p|^2}{\Delta_p^2+\gamma^2}+\frac{|\Omega_b|^2}{(\Delta+\Delta_p)^2+\gamma^2}\right)$ and long-lived state population $n_{ll}^{(2)}=\frac{2\Gamma_{nr}\gamma}{\Gamma_{ll}(\Gamma_r+\Gamma_{nr})}\left(\frac{|\Omega_p|^2}{\Delta_p^2+\gamma^2}+\frac{|\Omega_b|^2}{(\Delta+\Delta_p)^2+\gamma^2}\right)$. The susceptibility of the probe field is

$$\chi^{(3)}_{\text{cross-lin}}(\omega_b)=\alpha\frac{2\gamma}{\Delta+\Delta_p+i\gamma}\left(\frac{|\Omega_p|^2}{\Delta_p^2+\gamma_1^2}+\frac{|\Omega_b|^2}{(\Delta+\Delta_p)^2+\gamma^2}\right)\\+\frac{|\Omega_p|^2(\Delta+2i\gamma)}{(\Delta+\Delta_p+i\gamma)^2(\Delta_p-i\gamma)}\frac{1}{\Delta+i(\Gamma_r+2\Gamma_v+\Gamma_{nr})}, \quad (12)$$

with $\alpha\equiv\frac{\Gamma_{nr}}{\Gamma_{ll}(\Gamma_r+\Gamma_{nr})}+\frac{1}{\Gamma_r+\Gamma_{nr}}$.

We assume the homogeneous dephasing rate $\gamma_h$ is much larger than the various population decay rates, so $\gamma\gg\Gamma_r,\Gamma_v,\Gamma_{nr},\Gamma_{ll},\Delta$. Integrating over the inhomogeneous broadening, and using the relation $\int\frac{2i\gamma}{(\Delta_p+i\gamma)^2(\Delta_p-i\gamma)}d\Delta_p\approx\frac{\pi}{\gamma}$ when the inhomogeneous broadening is larger than $\gamma$, the results are summarized below.

*(a). Co-circular:*

$$\int d\Delta_p \chi^{(3)}_{\text{co-cir}}(\omega_b) \approx -\alpha \frac{i\pi}{\gamma}\left(|\Omega_p|^2 + |\Omega_b|^2\right)$$
$$+ \frac{\pi|\Omega_p|^2}{\gamma}\left(\frac{1}{\Delta + i(\Gamma_r + \Gamma_{nr})} + \frac{1}{\Delta + i(\Gamma_r + 2\Gamma_v + \Gamma_{nr})}\right) \quad (13)$$
$$+ \frac{\pi|\Omega_p|^2}{\gamma}\frac{i\Gamma_{nr}}{(\Delta + i\Gamma_{ll})(\Delta + i(\Gamma_r + \Gamma_{nr}))}.$$

Here $\alpha \equiv \frac{\Gamma_{nr}}{\Gamma_{ll}(\Gamma_r+\Gamma_{nr})} + \frac{1}{\Gamma_r+\Gamma_{nr}} + \frac{1}{\Gamma_r+2\Gamma_v+\Gamma_{nr}}$.

*(b). Cross-circular:*

$$\int d\Delta_p \chi^{(3)}_{\text{cross-cir}}(\omega_b) \approx -\frac{i\pi}{\gamma}\left(\alpha|\Omega_p|^2 + \beta|\Omega_b|^2\right). \quad (14)$$

Here $\alpha = \frac{\Gamma_{ll}+\Gamma_{nr}}{\Gamma_{ll}(\Gamma_r+\Gamma_{nr})} - \frac{1}{\Gamma_r+2\Gamma_v+\Gamma_{nr}}$ and $\beta = \frac{\Gamma_{ll}+\Gamma_{nr}}{\Gamma_{ll}(\Gamma_r+\Gamma_{nr})} + \frac{1}{\Gamma_r+2\Gamma_v+\Gamma_{nr}}$.

*(c). Co-linear:*

$$\int d\Delta_p \chi^{(3)}_{\text{co-lin}}(\omega_b) \approx -\alpha \frac{i\pi}{\gamma}\left(|\Omega_p|^2 + |\Omega_b|^2\right)$$
$$+ \frac{\pi|\Omega_p|^2}{\gamma}\frac{1}{\Delta + i(\Gamma_r + \Gamma_{nr})} \quad (15)$$
$$+ \frac{\pi|\Omega_p|^2}{\gamma}\frac{i\Gamma_{nr}}{(\Delta + i\Gamma_{ll})(\Delta + i(\Gamma_r + \Gamma_{nr}))}.$$

Here $\alpha \equiv \frac{\Gamma_{nr}}{\Gamma_{ll}(\Gamma_r+\Gamma_{nr})} + \frac{1}{\Gamma_r+\Gamma_{nr}}$.

*(d). Cross-linear:*

$$\int d\Delta_p \chi^{(3)}_{\text{cross-lin}}(\omega_b) \approx -\alpha \frac{i\pi}{\gamma}\left(|\Omega_p|^2 + |\Omega_b|^2\right)$$
$$+ \frac{\pi|\Omega_p|^2}{\gamma}\frac{1}{\Delta + i(\Gamma_r + 2\Gamma_v + \Gamma_{nr})}. \quad (16)$$

Here $\alpha \equiv \frac{\Gamma_{nr}}{\Gamma_{ll}(\Gamma_r+\Gamma_{nr})} + \frac{1}{\Gamma_r+\Gamma_{nr}}$.

Note that the co-circular and co-linear results both have three terms. The first term corresponds to the incoherent spectral hole-burning which appears as a broad background with a spectral width of $2\gamma$; the second term shows a narrow resonance which comes from the population pulsation of the excitons; while the third term is an ultra-narrow resonance induced by the population pulsation of the long-lived state (note that $\frac{i\Gamma_{nr}}{(\Delta+i\Gamma_{ll})(\Delta+i(\Gamma_r+\Gamma_{nr}))} \propto \frac{1}{\Delta+i\Gamma_{ll}} - \frac{1}{\Delta+i(\Gamma_r+\Gamma_{nr})}$). In the cross-linear case, there is no pulsation for the long-lived state so only the first and second terms remain. For the cross-circular case there is only the incoherent spectral hole-burning term.

If the inhomogeneous broadening is comparable to or smaller than $\gamma$, then we only need to quantitatively modify the background value and the coefficient before the resonance. The width of the population pulsation resonance is not affected.

The above results show how a pump laser affects the probe signal, as a function of probe-pump detuning $\Delta \equiv \omega_b - \omega_p$. In the co-polarized pump and probe measurement, the two lasers are set at equal power and the total nonlinear DR signal detected is in fact the sum of the DR of the probe beam due to the presence of pump beam, and the DR of the pump beam due to the presence of the probe beam. The second contribution can be obtained by replacing $\Delta$ by $-\Delta$ in the first contribution which is given in the above equations (Eq. (13)-(16)), since pump and probe are symmetric. The total signal then corresponds to the $\Delta$-symmetric part of these equations, which is proportional to $\text{Im}\left[\frac{1}{\Delta+i\Gamma}\right]$. Here $\Gamma$ denotes various population decay rate.

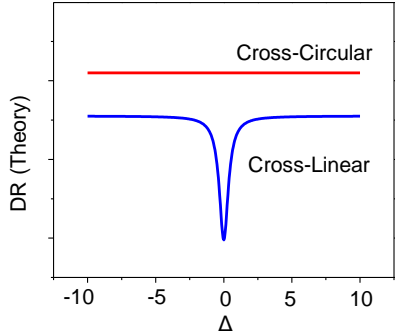

FIG. S6. Comparison of theoretical cross-linear and cross-circular DR responses. The DR response is calculated by taking the imaginary part of supplementary Eqns.14 and 16. The decay rates are chosen resemble the experimental values shown in Fig. 3(a). We assume that the intervalley scattering rate and exciton relaxation rates are equal and that non-radiative decay is negligible: $\gamma = 1000, \Gamma_r = 0.133, \Gamma_v = 0.133$, and $\Gamma_{nr} = 0$.

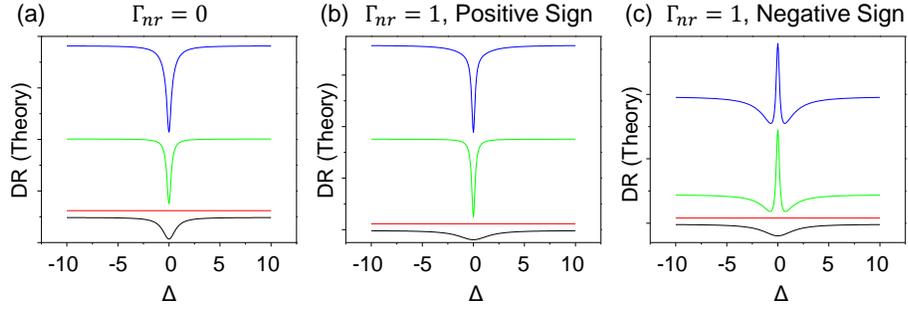

FIG. S7. Comparison of theoretical all four combinations of cross-linear (black) cross-circular (red), co-linear (green), and co-circular (blue) DR responses. The DR response is calculated by taking the imaginary part of supplementary Eqns. 13-16. (a) Plotting Eqns. 13-16 with: $\gamma = 1000, \Gamma_r = 0.25, \Gamma_v = 0.25$, and $\Gamma_{nr} = 0$, (b) using: $\gamma = 1000, \Gamma_r = 0.25, \Gamma_v = 0.25, \Gamma_{nr} = 1$ and $\Gamma_{ll} = 0.2$, without flipping the sign of the narrow resonance, (c) or using: $\gamma = 1000, \Gamma_r = 0.25, \Gamma_v = 0.25, \Gamma_{nr} = 1$ and $\Gamma_{ll} = 0.2$, with flipping the sign of the narrow resonance as discussed in Supplementary Note 4.

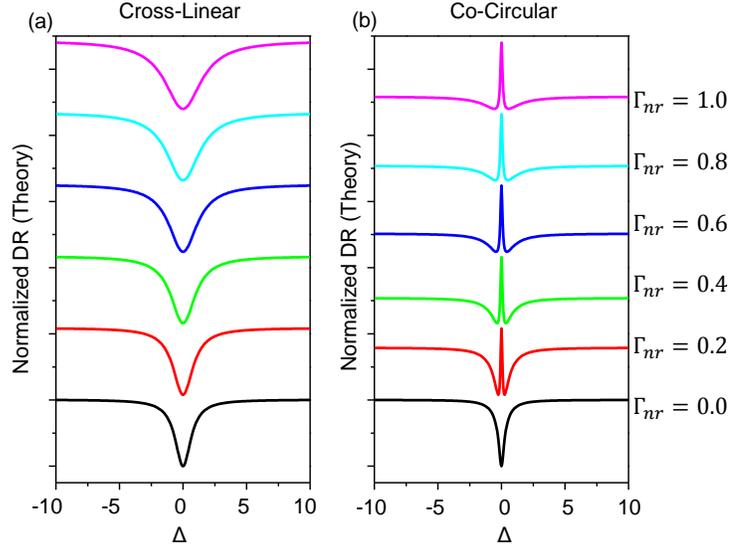

FIG. S8. Comparison of theoretical cross-linear (a) and co-circular (b) DR responses with the narrow resonance sign flipped (see Supplementary Note 4) for increasing values of $\Gamma_{nr}$, using: $\gamma = 1000, \Gamma_r = 0.25, \Gamma_v = 0.25$ and $\Gamma_{ll} = 0.2$. We assume that the intervalley scattering rate ($\Gamma_v$) is equal to the exciton relaxation rate ($\Gamma_r$). The difference between linewidths of the dip features is determined by the relative values of $\Gamma_r$ to $\Gamma_v$. The range of $\Gamma_{nr}$ plotted is chosen to simulate the broadening observed in cross linear spectra of Fig. 4(a) of the main text.

**Supplementary Note 2: Comparison to the excitonic picture**

In a more rigorous sense, the exciton should be treated as a bound state of an electron-hole pair with the excitonic operator given by $\hat{X}^\dagger_{\mathbf{k},\pm} \equiv \sum_{\mathbf{q}} \Phi_{\mathbf{k},\pm}(\mathbf{q}) \hat{e}^\dagger_{\pm \mathbf{K}+\frac{\mathbf{k}}{2}+\mathbf{q}} \hat{h}^\dagger_{\pm \mathbf{K}+\frac{\mathbf{k}}{2}-\mathbf{q}}$. Here $\mathbf{k}$ is the exciton center-of-mass wave vector, $\hat{e}^\dagger_{\pm \mathbf{K}+\mathbf{q}}$ ($\hat{h}^\dagger_{\pm \mathbf{K}+\mathbf{q}}$) is the creation operator for a Bloch electron (hole) in the $\pm K$ valley, and $\Phi_{\mathbf{k},\pm}(\mathbf{q})$ is the wave function of their relative motion. Because the exciton wave function extension in $\mathbf{k}$-space is $\sim 1/a_B$, which is two orders of magnitude larger than the light cone size, the excitons inside and outside the light cone should have rather similar wave functions $\Phi_{\mathbf{k},\pm}(\mathbf{q})$. Below we using $\hat{X}_\pm$ to denote all excitons both inside and outside the light cone. In the low density limit, the coherent dynamics of the exciton polarization $P_\pm \equiv \langle \hat{X}_\pm \rangle$ and population $n_\pm \equiv \langle \hat{X}^\dagger_\pm \hat{X}_\pm \rangle$ can be described by the excitonic Bloch equations [4]:

$$i\frac{dP_\pm}{dt}\bigg|_c = (\omega_X + u n_\pm + u_D n_D) P_\pm - (1 - f n_\pm - f_{ll} n_{ll}) \Omega_\pm(t),$$
$$\frac{dn_\pm}{dt}\bigg|_c = -2(1 - f n_\pm - f_{ll} n_{ll}) \text{Im}[\Omega_\pm(t) P^*_\pm]. \tag{17}$$

Here, the $f n_\pm$ and $u n_\pm$ terms account for the phase-space-filling of the excitons and interaction between the excitons respectively. We also added $f_{ll} n_{ll}$ and $u_{ll} n_{ll}$ to phenomenologically account for the phase-space-filling from the long-lived state and the interaction between the long-lived state and the excitons, respectively. The relative strengths of these effects can be tuned by the parameters, $f_{ll}$ and $u_{ll}$.

The incoherent processes are the same as in Eq. (6) in Supplementary Note 1. Again we make an expansion in terms of $|\Omega_\pm|$, and write $P_\pm = P^{(1)}_\pm + P^{(3)}_\pm + \cdots$, $n_\pm = n^{(2)}_\pm + n^{(4)}_\pm + \cdots$, $n_D = n^{(2)}_D + n^{(4)}_D \cdots$, $N \equiv n_+ + n_-$ and $\Delta N \equiv n_+ - n_-$. Up to the third order,

$$i\frac{dP_\pm^{(1)}}{dt} = (\omega_X - i\gamma)P_\pm^{(1)} - \Omega_\pm(t),$$

$$\frac{dn_\pm^{(2)}}{dt} = -2\text{Im}\left[\Omega_\pm \left(P_\pm^{(1)}\right)^*\right] - (\Gamma_r + \Gamma_v + \Gamma_{nr})n_\pm^{(2)} + \Gamma_v n_\mp^{(2)},$$

$$\frac{dn_{ll}^{(2)}}{dt} = -\Gamma_{ll}n_{ll}^{(2)} + \Gamma_{nr}\left(n_+^{(2)} + n_-^{(2)}\right), \quad (18)$$

$$i\frac{dP_+^{(3)}}{dt} = (\omega_X - i\gamma)P_+^{(3)} + \left(un_+^{(2)} + u_{ll}n_{ll}^{(2)}\right)P_+^{(1)} + \left(fn_+^{(2)} + f_{ll}n_{ll}^{(2)}\right)\Omega_+,$$

$$i\frac{dP_-^{(3)}}{dt} = (\omega_X - i\gamma)P_-^{(3)} + \left(un_-^{(2)} + u_{ll}n_{ll}^{(2)}\right)P_-^{(1)} + \left(fn_+^{(2)} + f_{ll}n_{ll}^{(2)}\right)\Omega_-.$$

Below we derive the susceptibility for co-circular pump-probe. In this case $\Omega_+ = \Omega_p e^{-i\omega_p t} + \Omega_b e^{-i\omega_b t}$ and $\Omega_- = 0$, so $P_+^{(1)} = -\frac{\Omega_p e^{-i\omega_p t}}{\Delta_p + i\gamma} - \frac{\Omega_b e^{-i\omega_b t}}{\Delta + \Delta_p + i\gamma}$, $P_-^{(1)} = 0$, $N^{(2)} = \frac{2\gamma}{\Gamma_r + \Gamma_{nr}}\left(\frac{|\Omega_p|^2}{\Delta_p^2 + \gamma^2} + \frac{|\Omega_b|^2}{(\Delta + \Delta_p)^2 + \gamma^2}\right) + \left(Ae^{-i\Delta t} + \text{c.c.}\right)$, $\Delta N^{(2)} = \frac{2\gamma}{\Gamma_r + 2\Gamma_v + \Gamma_{nr}}\left(\frac{|\Omega_p|^2}{\Delta_p^2 + \gamma^2} + \frac{|\Omega_b|^2}{(\Delta + \Delta_p)^2 + \gamma^2}\right) + \left(Be^{-i\Delta t} + \text{c.c.}\right)$, $n_{ll}^{(2)} = \frac{2\Gamma_{nr}\gamma}{\Gamma_{ll}(\Gamma_r + \Gamma_{nr})}\left(\frac{|\Omega_p|^2}{\Delta_p^2 + \gamma^2} + \frac{|\Omega_b|^2}{(\Delta + \Delta_p)^2 + \gamma^2}\right) + \left(\frac{i\Gamma_{nr}}{\Delta + i\Gamma_{ll}}Ae^{-i\Delta t} + \text{c.c.}\right)$ with $A = \frac{\Omega_b \Omega_p^*}{\Delta + i(\Gamma_r + \Gamma_{nr})}\frac{\Delta + 2i\gamma}{(\Delta_p - i\gamma)(\Delta + \Delta_p + i\gamma)}$ and $B = \frac{\Omega_b \Omega_p^*}{\Delta + i(\Gamma_r + 2\Gamma_v + \Gamma_{nr})}\frac{\Delta + 2i\gamma}{(\Delta_p - i\gamma)(\Delta + \Delta_p + i\gamma)}$. The optical susceptibility of the probe field is,

$$\chi_{\text{co-cir}}^{(3)}(\omega_b) = \left(\frac{\alpha\gamma}{\Delta + \Delta_p + i\gamma} - \frac{\beta\gamma}{(\Delta + \Delta_p + i\gamma)^2}\right)\left(\frac{|\Omega_p|^2}{\Delta_p^2 + \gamma^2} + \frac{|\Omega_b|^2}{(\Delta + \Delta_p)^2 + \gamma^2}\right)$$
$$+ \frac{|\Omega_p|^2(\Delta + 2i\gamma)}{(\Delta + \Delta_p + i\gamma)^2(\Delta_p - i\gamma)}\left(\frac{\frac{f}{2} + \frac{i\Gamma_{nr}}{\Delta + i\Gamma_D}f_D}{\Delta + i(\Gamma_r + \Gamma_{nr})} + \frac{\frac{f}{2}}{\Delta + i(\Gamma_r + 2\Gamma_v + \Gamma_{nr})}\right) \quad (19)$$
$$- \frac{|\Omega_p|^2(\Delta + 2i\gamma)}{(\Delta + \Delta_p + i\gamma)^2(\Delta_p^2 + \gamma^2)}\left(\frac{\frac{u}{2} + \frac{i\Gamma_{nr}}{\Delta + i\Gamma_D}u_D}{\Delta + i(\Gamma_r + \Gamma_{nr})} + \frac{\frac{u}{2}}{\Delta + i(\Gamma_r + 2\Gamma_v + \Gamma_{nr})}\right),$$

where $\alpha \equiv \frac{2f_D\Gamma_{nr}}{\Gamma_{ll}(\Gamma_r + \Gamma_{nr})} + \frac{f}{\Gamma_r + \Gamma_{nr}} + \frac{f}{\Gamma_r + 2\Gamma_v + \Gamma_{nr}}$, $\beta \equiv \frac{2u_D\Gamma_{nr}}{\Gamma_{ll}(\Gamma_r + \Gamma_{nr})} + \frac{u}{\Gamma_r + \Gamma_{nr}} + \frac{u}{\Gamma_r + 2\Gamma_v + \Gamma_{nr}}$.

Integrating over a large inhomogeneous broadening, we end up with a simpler equation,

$$\int d\Delta_p \chi_{\text{co-cir}}^{(3)}(\omega_b) \approx \frac{\pi}{2i\gamma}\left(\alpha + i\frac{\beta}{2\gamma}\right)\left(|\Omega_p|^2 + |\Omega_b|^2\right)$$
$$+ \frac{\pi|\Omega_p|^2}{2\gamma}\left(f + i\frac{u}{2\gamma}\right)\left(\frac{1}{\Delta + i(\Gamma_r + \Gamma_{nr})} + \frac{1}{\Delta + i(\Gamma_r + 2\Gamma_v + \Gamma_{nr})}\right) \quad (20)$$
$$+ \frac{\pi|\Omega_p|^2}{\gamma}\left(f_{ll} + i\frac{u_{ll}}{2\gamma}\right)\frac{i\Gamma_{nr}}{(\Delta + i\Gamma_{ll})(\Delta + i(\Gamma_r + \Gamma_{nr}))},$$

where we have assumed $\gamma$ is large compared to the range of pump-probe detuning, $\Delta$.

Compared to the previous result from the optical Bloch equations of two-level systems (Eq. (13) in Supplementary Note 1), we can see that the only difference is the constant factors, $\alpha + i\frac{\beta}{2\gamma}$, $f + i\frac{u}{2\gamma}$, and $f_{ll} + i\frac{u_{ll}}{2\gamma}$, that account for the phase space filling and resonance shift by the interactions. Note that these factors are complex which can change the relative weight of $\text{Re}\left(\chi^{(3)}(\omega)\right)$ and $\text{Im}\left(\chi^{(3)}(\omega)\right)$. Nevertheless, since $\text{Re}\left(\chi^{(3)}(\omega)\right)$ and $\text{Im}\left(\chi^{(3)}(\omega)\right)$ have comparable linewidths determined by the population decay rates, this relative weight is not that crucial in extracting the decay rates.

The results for the other three polarization combinations can be obtained similarly. Their expressions are not shown here, since the difference from the results Eq. (14)-(16) in Supplementary Note 1 is the constant factors given above.

**Supplementary Note 3: Phenomenological analysis of excitons outside the light cone.**

In order for a better understanding to those excitons outside the light cone, we consider the model shown in Fig. S9. Here for simplicity we only consider +K valley and ignore the valley relaxation. $|+\rangle$ is the bright exciton inside the light cone which has an intrinsic radiative decay rate $\Gamma_{IR}$, and $|D+\rangle$ is the exciton outside the light cone. Since the light cone edge is two orders of magnitude smaller than $1/a_B$, we expect $|+\rangle$ and $|D+\rangle$ to have similar wave functions, comparable energy, and can interconvert with each other through scattering. Both $|+\rangle$ and $|D+\rangle$ can relax to the lower energy long-lived state $|ll\rangle$ with rate $\Gamma_{nr}$, but we assume $|ll\rangle$ can't scatter back to $|+\rangle$ or $|D+\rangle$. $|0\rangle$ is the vacuum state with zero exciton. The relaxation channels and the corresponding rates are illustrated in the Fig. S9.

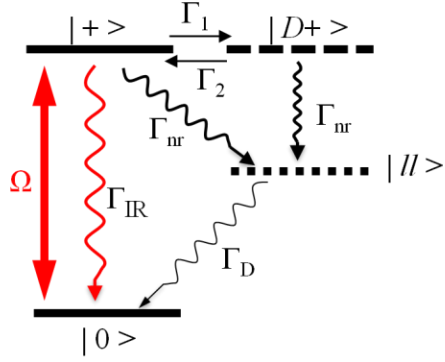

FIG. S9. The model with exciton inside the light cone $|+\rangle$, outside the light cone $|D+\rangle$, low energy long-lived state $|ll\rangle$, and vacuum state $|0\rangle$. Their relaxation channels and rates are illustrated with arrows.

When the system is driven by co-circular pump and probe lasers, the pulsation part of the exciton population satisfy the rate equations:

$$\frac{dn_+}{dt} = \Gamma_2 n_{D+} - (\Gamma_1 + \Gamma_{nr} + \Gamma_{IR})n_+ + (\Omega e^{-i\Delta t} + \text{c.c.}),$$
$$\frac{dn_{D+}}{dt} = \Gamma_1 n_+ - (\Gamma_2 + \Gamma_{nr})n_{D+}, \qquad (21)$$
$$\frac{dn_{ll}}{dt} = \Gamma_{nr}(n_+ + n_{D+}) - \Gamma_{ll} n_{ll}.$$

The stationary solution is given by,

$$n_+ = \frac{i(\Delta + i(\Gamma_2 + \Gamma_{nr}))\Omega e^{-i\Delta t}}{(\Delta + i(\Gamma_1 + \Gamma_{nr} + \Gamma_{IR}))(\Delta + i(\Gamma_2 + \Gamma_{nr})) + \Gamma_1 \Gamma_2} + \text{c.c.},$$
$$n_{D+} = \frac{-\Gamma_1 \Omega e^{-i\Delta t}}{(\Delta + i(\Gamma_1 + \Gamma_{nr} + \Gamma_{IR}))(\Delta + i(\Gamma_2 + \Gamma_{nr})) + \Gamma_1 \Gamma_2} + \text{c.c.}, \qquad (22)$$
$$n_{ll} = \frac{-\Gamma_{nr}}{\Delta + i\Gamma_{ll}} \cdot \frac{(\Delta + i(\Gamma_1 + \Gamma_2 + \Gamma_{nr}))\Omega e^{-i\Delta t}}{(\Delta + i(\Gamma_1 + \Gamma_{nr} + \Gamma_{IR}))(\Delta + i(\Gamma_2 + \Gamma_{nr})) + \Gamma_1 \Gamma_2} + \text{c.c.}.$$

Under a low exciton density we expect the exciton scattering rate is small, $\Gamma_{IR} \gg \Gamma_1$, we have $(\Gamma_1 + \Gamma_{IR})\Gamma_2 \gg \Gamma_1 \Gamma_2$ thus $(\Delta + i(\Gamma_1 + \Gamma_{nr} + \Gamma_{IR}))(\Delta + i(\Gamma_2 + \Gamma_{nr})) + \Gamma_1 \Gamma_2 \approx (\Delta + i(\Gamma_1 + \Gamma_{nr} + \Gamma_{IR}))(\Delta + i(\Gamma_2 + \Gamma_{nr}))$, the above solution simplifies to

$$n_+ \approx \frac{i\Omega e^{-i\Delta t}}{\Delta + i(\Gamma_1 + \Gamma_{nr} + \Gamma_{IR})} + \text{c.c.},$$
$$n_{D+} \approx \frac{-\Gamma_1 \Omega e^{-i\Delta t}}{(\Delta + i(\Gamma_1 + \Gamma_{nr} + \Gamma_{IR}))(\Delta + i(\Gamma_2 + \Gamma_{nr}))} + \text{c.c.}, \qquad (23)$$
$$n_{ll} \approx \frac{-\Gamma_{nr}}{\Delta + i\Gamma_{ll}} \cdot \frac{(\Delta + i(\Gamma_1 + \Gamma_2 + \Gamma_{nr}))\Omega e^{-i\Delta t}}{(\Delta + i(\Gamma_1 + \Gamma_{nr} + \Gamma_{IR}))(\Delta + i(\Gamma_2 + \Gamma_{nr}))} + \text{c.c.}.$$

So $n_+$ contributes to a resonance with width $\Gamma_1 + \Gamma_{nr} + \Gamma_{IR}$. $n_{D+}$ has two resonances with width $\Gamma_1 + \Gamma_{nr} + \Gamma_{IR}$ and $\Gamma_2 + \Gamma_{nr}$. $n_{ll}$ has three resonances with width $\Gamma_1 + \Gamma_{nr} + \Gamma_{IR}$, $\Gamma_2 + \Gamma_{nr}$ and $\Gamma_{ll}$.

Under a moderate or high exciton density, we expect the exciton-exciton scattering rate is large so $\Gamma_1 + \Gamma_2 \gg \Gamma_{IR}$. In this case $\big(\Delta + i(\Gamma_1 + \Gamma_{nr} + \Gamma_{IR})\big)\big(\Delta + i(\Gamma_2 + \Gamma_{nr})\big) + \Gamma_1\Gamma_2 \approx \big(\Delta + i(\Gamma_1 + \Gamma_2 + \Gamma_{nr} + \Gamma_{IR})\big)\big(\Delta + i\big(\frac{\Gamma_2}{\Gamma_1+\Gamma_2}\Gamma_{IR} + \Gamma_{nr}\big)\big)$, so

$$n_+ \approx \frac{i\big(\Delta + i(\Gamma_2 + \Gamma_{nr})\big)\Omega e^{-i\Delta t}}{\big(\Delta + i(\Gamma_1 + \Gamma_2 + \Gamma_{nr} + \Gamma_{IR})\big)\big(\Delta + i\big(\frac{\Gamma_2}{\Gamma_1+\Gamma_2}\Gamma_{IR} + \Gamma_{nr}\big)\big)} + \text{c.c.},$$

$$n_{D+} \approx \frac{-\Gamma_1\Omega e^{-i\Delta t}}{\big(\Delta + i(\Gamma_1 + \Gamma_2 + \Gamma_{nr} + \Gamma_{IR})\big)\big(\Delta + i\big(\frac{\Gamma_2}{\Gamma_1+\Gamma_2}\Gamma_{IR} + \Gamma_{nr}\big)\big)} + \text{c.c.}, \quad (24)$$

$$n_{ll} \approx \frac{-\Gamma_{nr}\Omega e^{-i\Delta t}}{(\Delta + i\Gamma_{ll})\big(\Delta + i\big(\frac{\Gamma_2}{\Gamma_1+\Gamma_2}\Gamma_{IR} + \Gamma_{nr}\big)\big)} + \text{c.c.}.$$

Now we get three resonances with width $\Gamma_1 + \Gamma_2 + \Gamma_{nr} + \Gamma_{IR}$, $\frac{\Gamma_2}{\Gamma_1+\Gamma_2}\Gamma_{IR} + \Gamma_{nr}$ and $\Gamma_{ll}$, respectively. Considering the detailed balance between $|+\rangle$ and $|D+\rangle$ which requires $n_+\Gamma_1 = n_{D+}\Gamma_2$, so the second resonance width $\frac{\Gamma_2}{\Gamma_1+\Gamma_2}\Gamma_{IR} + \Gamma_{nr} = \frac{n_+}{n_++n_{D+}}\Gamma_{IR} + \Gamma_{nr}$ corresponds to the average decay rate of the all excitons both inside and outside the light cone.

We note that $\Gamma_1 + \Gamma_{nr} + \Gamma_{IR}$ is the total decay rate of the bright exciton $|+\rangle$, which is expected to be much larger than our experimental frequency detuning range. Thus in the above two cases, the largest resonance width ($\Gamma_1 + \Gamma_{nr} + \Gamma_{IR}$ or $\Gamma_1 + \Gamma_2 + \Gamma_{nr} + \Gamma_{IR}$) is not detected. The experimentally measured long timescale > 6 ns should correspond to the long-lived state lifetime $\Gamma_{ll}$. The ~1.7 ns lifetime could correspond to $\Gamma_2 + \Gamma_{nr}$ in the weak scattering case, or $\frac{n_+}{n_++n_{D+}}\Gamma_{IR} + \Gamma_{nr}$ in the moderate or strong scattering case. We expect that the latter is more probable because the extremely large Coulomb interaction in TMD monolayer favors a high exciton-exciton scattering rate, then $\Gamma_r = \frac{n_+}{n_++n_{D+}}\Gamma_{IR}$ accounts for the overall radiative decay of exciton population inside and outside the light cone.

**Supplementary Note 4: Possible origin for the sign change of the ultra-narrow resonance**

The probe differential reflectance signal is proportional to $A \cdot \text{Re}(\chi^{(3)}) + B \cdot \text{Im}(\chi^{(3)})$. From the $\chi^{(3)}$ expression given in Eq. (20) for the co-circular case, the signal which accounts for the observed resonances is proportional to

$$\left(Bf - A\frac{u}{2\gamma}\right)\text{Im}\left(\frac{1}{\Delta + i(\Gamma_r + \Gamma_{nr})} + \frac{1}{\Delta + i(\Gamma_r + 2\Gamma_v + \Gamma_{nr})}\right) \\ + 2\left(Bf_{ll} - A\frac{u_{ll}}{2\gamma}\right)\text{Im}\left(\frac{i\Gamma_{nr}}{(\Delta + i\Gamma_{ll})(\Delta + i(\Gamma_r + \Gamma_{nr}))}\right). \tag{25}$$

In the above equation, we only kept the $\Delta$-symmetric terms since pump and probe signals are simultaneously detected. The differences between the phase-space-filling parameters $f$ and $f_{ll}$, manybody interaction parameters $\frac{u}{2\gamma}$ and $\frac{u_{ll}}{2\gamma}$ can lead to opposite signs for the two coefficients $Bf - A\frac{u}{2\gamma}$ and $Bf_{ll} - A\frac{u_{ll}}{2\gamma}$. In this case, the resonance from exciton population pulsation and the other resonance from long-lived state population pulsation will correspond to a peak and a dip respectively.

**Supplementary Note 5: Additional Methods**

Bulk MoSe$_2$ crystal was grown by vapor transport as described in previous work [5]. Optical studies presented in the main text were performed in vacuum on a single MoSe$_2$ monolayer exfoliated onto a SiO$_2$ substrate, held at 30 K in a liquid helium cold finger cryostat. Photoluminescence measurements were performed using a single grating spectrometer and cooled CCD camera. For the photoluminescence measurements a 20 µW, 532 nm laser off-resonantly excited the system. In the DR measurements, (peak) pump and probe powers were 2-80 µW with a duty cycle of 50%. For DR measurements, the pump and probe were (square-wave) amplitude modulated (~700 kHz) with acousto-optic modulators whose drivers were phase locked. The difference frequency between these modulation frequencies was the reference for a lock-in amplifier. The laser polarizations were adjusted with suitable combinations of achromatic half and quarter wave plates and polarizers. For the higher power DR measurements (Fig. 2 and Fig. 4), the reflection was detected with an amplified silicon photodiode. For the low power measurements (Fig. 3), a silicon APD and current preamplifier was used. For the degenerate DR measurements, a single narrow bandwidth Ti:sapphire laser was split into pump and probe beams, and a small detuning (~1 µeV) is applied to the pump and probe by the acousto-optic modulators. For the non-degenerate measurements, two tunable lasers (< 4 neV spectral width) were used.